\documentclass[aip,apl,reprint]{revtex4-1}
\usepackage{amsmath}
 \usepackage{graphicx}

\begin{document}
\title{Graphene p-n junctions with nonuniform Rashba spin-orbit coupling}
\author{Marek Rataj}
\affiliation{Faculty of Physics, Adam Mickiewicz University,
Umultowska 85, 61-614 Pozna\'n, Poland}
\affiliation{The Nano-Bio-Medical Centre, Umultowska 85, 61-614 Pozna\'n, Poland}
\author{J\'ozef Barna\'s}
\affiliation{Faculty of Physics, Adam Mickiewicz University,
Umultowska 85, 61-614 Pozna\'n, Poland}

\begin{abstract}
Linear conductance of graphene-based \mbox{$p$-$n$} junctions with
Rashba spin-orbit coupling is considered theoretically. A~square potential step is used to model the~junctions, while the~coupling is introduced in terms of the~Kane-Mele model (C.L. Kane and E.J. Mele, Phys.Rev.Lett. 95, 226801(2005)). The
main objective is a~description of electronic transport in
junctions where Rashba parameter is nonuniform. Such a~nonuniformity can appear when graphene is asymmetrically covered
with atomic layers, or when Rashba coupling is strongly
dependent on electric field. It is shown that conductance is significantly modified by the~considered nonuniformity, which is most clearly manifested by an~anomalous minimum at a~certain potential step height.
\end{abstract}
\maketitle

Owing to the electric field effect~\cite{Novoselov2004}, various
all-graphene-based heterostructures can be easily made by means of
electrostatic gates. The simplest structures of this kind are
\mbox{$p$-$n$} junctions, which have been already analysed
theoretically\cite{Katsnelson2006}
and investigated
experimentally\cite{Lemme2007}. Generally,
transport characteristics of such junctions significantly depend
on the spin-orbit interactions in
graphene\cite{Yamakage2009,Yamakage2011,Bai2010a}. This is mainly due to the fact that according to the
commonly used Kane-Mele model\cite{Kane2005}, the intrinsic coupling opens a~band gap, while the Rashba coupling (RSOC)
splits both the conduction and valence bands. In this paper we focus on the latter, motivated by recent results showing that it can be made significantly dominant in graphene, as pointed out in Refs. \onlinecite{Varykhalov2008,Sanchez-Barriga2010,Rashba2009,Li2011,CastroNeto2009,Ertler2009,Dedkov2008}, where RSOC parameters of the order of $10$ meV were reported.

The parameter of RSOC has been so far treated in transport investigations as a~constant throughout the structure and also independent of the gate voltages used to make \mbox{$p$-$n$}
junctions\cite{Yamakage2009,Yamakage2011}. However, this parameter
can be generally nonuniform, which may occur when graphene is covered with an atomic layer from
bottom in one part of the junction and on top in the second part~\cite{Varykhalov2008,Sanchez-Barriga2010,Rashba2009,Li2011,CastroNeto2009,Ertler2009,Dedkov2008}, or when RSOC is strongly dependent on electric
field\cite{Huertas-Hernando2006,Min2006,Gmitra2009}.
This problem is addressed in the present paper, where we show that
conductance of such \mbox{$p$-$n$} junctions differs from that of
junctions with constant spin-orbit coupling. We
focus especially on a situation in which the RSOC parameter changes sign at
the border between the two parts of the junction. All this
gives rise to anomalous behaviour of the corresponding conductance.

In order to describe charge carriers in graphene we employ the
effective low-energy Hamiltonian\cite{Wallace1947}, $H_0 = -i
\hbar v_{F} (\sigma_x
\partial_x + \sigma_y \partial_y ) \otimes s_0$, where $v_F$ is the
electron velocity. We use the notation with $\sigma_\alpha$ and
$s_\alpha$ being the Pauli matrices (for $\alpha=x,y,z$) in the
pseudospin and spin spaces, respectively, and $\sigma_0$ and $s_0$
denoting the corresponding unit matrices. We restrict our
considerations to a~single electronic valley, neglecting thus any
inter-valley scattering. RSOC is
taken in the form introduced by Kane and Mele\cite{Kane2005}, $
H_{R} = \lambda (\sigma_{x} \otimes s_{y} - \sigma_{y} \otimes
s_{x})$, where $\lambda$ is the coupling parameter. The full
Hamiltonian, $H=H_0+H_R$, has the following eigenvalues:
\begin{equation}
\label{energy} E = l \lambda + s \sqrt{\lambda^2 + (k_x^2+k_y^2)
v_F^2 \hbar^2},
\end{equation}
where $\boldsymbol{k}$ is the two-dimensional wave vector, $s=1$
($s=-1$) for the conduction (valence) band, and $l=\pm 1$ is used
to distinguish between the two subbands.

The \mbox{$p$-$n$} junction is described by the term, $V(x,y)=V_0
\Theta(x) \sigma_0 \otimes s_0$, to be added to the full
Hamiltonian. Let us denote the Fermi energy (measured from the
neutrality point in the left part of the junction) by $E_F$.

Assume now a~particle of energy $E$ at the Fermi level, $E=E_F$,
in the subband $l$ is incident on the potential step at an
arbitrary angle $\phi$. The corresponding wave function has the
form of a~plane wave multiplied by a~column vector derived in
Ref.~\onlinecite{Yamakage2009}, $\psi^i_{E,\phi,l}(x,y)=\left\vert
s k_x,E,\phi,l \right\rangle \exp[ i s  k_x x]  \exp[ i k_y
y]\equiv\psi^i_{E,\phi,l}(x)\exp[ i k_y y]$, where $k_x$ is a
positive solution of Eq.~(\ref{energy}) (note $s$ ensures that
group velocity points in the right direction). In the same way one
can build the wave functions for carriers reflected and
transmitted into the subband $l$; $r_l\psi^r_{E,\phi,l}(x,y)$ and
$t_l\psi^t_{E-V_0,\phi,l}(x,y)$, and for carriers reflected and
transmitted into the other subband, $l^\prime\ne l$;
$r_l^\prime\psi^r_{E,\phi,l'}(x,y)$ and
$t_l^\prime\psi^t_{E-V_0,\phi,l'}(x,y)$. Here, $r_l$, $t_l$,
$r'_l$ and $t'_l$ are the corresponding reflection and
transmission amplitudes. Owing to the translational symmetry along
the $y$ direction, the $k_y$ component of the wave vector is
conserved.

To find the linear conductance we need to know the amplitudes
$t_l$ and $t_l'$, which can be found from the continuity condition
at the boundary $x=0$,
\begin{equation}
\begin{split}
&\psi^i_{E,\phi,l}(x=0) + r_l\psi^r_{E,\phi,l}(x=0) + r'_l\psi^r_{E,\phi,l'}(x=0) \\
&= t_l \psi^t_{E-V_0,\phi,l}(x=0) +t'_l
\psi^t_{E-V_0,\phi,l'}(x=0).
\end{split}
\end{equation}
Having found the transmission amplitudes and taking also into
account the respective group velocities, one obtains the
corresponding transmission probabilities $T_l$ and $T'_l$. The
full linear conductance can be then obtained by integrating over
all incidence angles,
\begin{equation}
G=\frac{2 k_F e^2 W}{\pi h} \int d\phi \sum_{l}\frac{k_{F}^{(l)}}{k_F}(T_l+T'_l)
\cos(\phi).
\end{equation}
In the following, the conductance will be normalized to the
factor $G_0$ defined as $G_0=2 e^2 W k_F / (h \pi)$, where $k_F$  is an average
of incident Fermi wave vectors $k_F^{(l)}$, while $e$ is the electron
charge and $W$ is the sample width.

\begin{figure}
\includegraphics{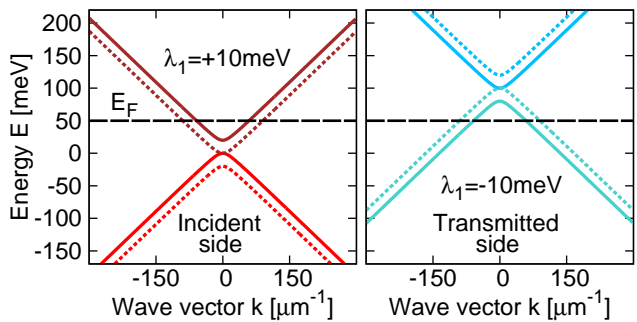}
\caption{Band structure of graphene for $\vert\lambda_1\vert=10$
meV, $\alpha^{-1}=0$, $E_F=50$ meV, and  $V_0=100$ meV.  Left part
shows the spectrum on the incident side of the junction
(where $\lambda_1=10$ meV), while the right part shows the
spectrum on the transmitted side (where $\lambda_1=-10$
meV). The solid (dotted) lines distinguish subband with opposite index $l$. \label{bands1}}
\end{figure}

The RSOC parameter $\lambda$ can in general be separated into
two terms; $\lambda=\lambda_1 + \lambda_2(\boldsymbol{E})$. The
first term is independent of electric field  -- it may be due to
the adjacent atomic layers (substrate/cover layers). The second
term can be controlled by external electric field $\boldsymbol{E}$
(gate voltages), $\lambda_2(\boldsymbol{E})=\alpha_1
\boldsymbol{z}\cdot\boldsymbol{E}$, where $\boldsymbol{z}$ is a
unit vector perpendicular to the graphene plane, and $\alpha_1$ is
a relevant parameter. The Fermi level in graphene is also
controlled by electric field, and this dependence can be expressed
as $\tilde{E}_F= {\rm sign}(\boldsymbol{z}\cdot\boldsymbol{E})
\sqrt{|\boldsymbol{z}\cdot\boldsymbol{E}|}\alpha_2 $, where
$\alpha_2$ is a~parameter, while $\tilde{E}_F=E_F$ for $x<0$ and
$\tilde{E}_F=E_F-V_0$ for $x>0$. Thus, when the Fermi level is
tuned by the gate voltage, the RSOC parameter $\lambda$ has to
be adjusted according to the formula
\begin{equation}
\lambda= \lambda_1  + {\rm sign}(\tilde{E}_F) \alpha^{-1}
\tilde{E}_F^2, \label{alpha}
\end{equation}
where $\alpha=\alpha_2^2/\alpha_1$. In the following we will
consider two options for the sign reversal of the RSOC parameter
at the potential step. First situation is when $\alpha^{-1}=0$ and
$\lambda_1$ is positive on one side of the step and negative on
the other side. The second possibility appears when $\alpha^{-1}$
is sufficiently large, while $\lambda_1$ is small (we take $\lambda_1=0$ for clarity of the argument)
and constant through the structure. We begin with the former.

\begin{figure}
\includegraphics{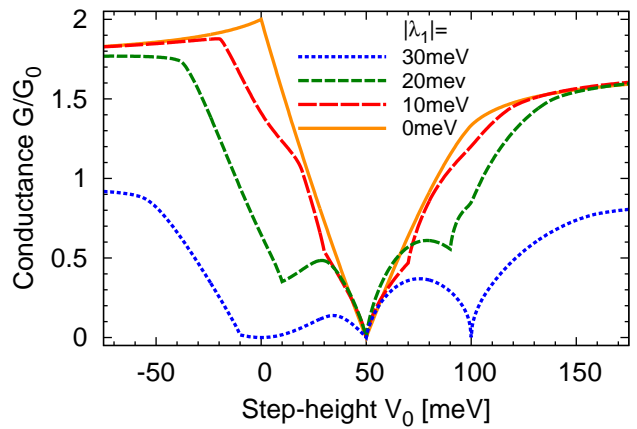}
\caption{Normalized conductance versus potential step height $V_0$
for $\vert\lambda_1\vert$ as indicated, with $\lambda_1$ positive
on the incidence side and negative on the transmitted side. The
other parameters are as in Fig.1}\label{Ge1}
\end{figure}

In Fig.~\ref{bands1} we show the band structure of graphene on the
two sides of the junction for $\vert\lambda_1\vert=10$ meV. The assumed
value is typical for graphene covered by atomic planes~\cite{Varykhalov2008,Sanchez-Barriga2010,Rashba2009,Li2011,CastroNeto2009,Ertler2009,Dedkov2008}. The parameter $\lambda_1$
is positive on the incident side and negative on the transmitted one (an opposite case would be fully symmetrical). Note,
that apart from an up-shift, the spectrum on the transmitted side is
inverted with respect to that on the incident side. This inversion is
due to the sign reversal of the RSOC parameter $\lambda_1$ at
the potential step.

Figure~\ref{Ge1} shows how the corresponding conductance $G$
depends on the potential step height $V_0$ for four different
values of $\vert\lambda_1\vert$. The curve corresponding to $\lambda_1=0$ serves as a reference. For the two curves corresponding to the lower values of $\vert\lambda_1\vert\neq 0$ the incident carriers come
from both conduction subbands (as in the case shown in
Fig.~\ref{bands1}). Therefore the corresponding conductance reaches values above $G/G_0=1$. There is a general trend in these curves that, when $V_0$ moves from the global minimum at $50$ meV towards larger values, the conductance grows until it saturates, similarly to the $\vert\lambda_1\vert = 0$ case. When $V_0$ moves towards lower values in turn, the conductance also grows at first and finally saturates, but the maximum present for $\lambda_1=0$ at $V_0=0$ is suppressed. This reduction is due to band mismatch created by inverted band splitting on both sides of the
junction. Moreover, in general Fermi level on the transmitted
side crosses either one or two subbands. The point, where a~new
subband enters (or leaves) the conduction regime is
associated with a~kink in the conductance curve. The curve for $\vert\lambda_1\vert=30$ meV differs from the others. First of all it reaches values only up to $G/G_0=1$, since the carriers on the incident side come from only one subband. Moreover, at $V_0=0$ the aforementioned band mismatch leads to a minimum equal zero. Furthermore, only one of the kinks mentioned above is visible in this case (at $V_0=-10$ meV). Finally, at $V_0=100$ meV an anomalous conductance minimum occurs.

\begin{figure}
\includegraphics{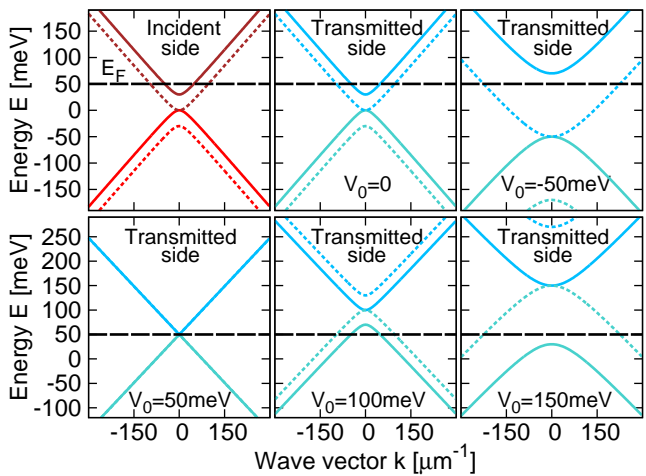}
\caption{Band structure of graphene for $\lambda_1=0$,
$\alpha=0.25$ eV, $E_F=50$ meV, and for indicated values of the
potential step height $V_0$.  Part (a) shows the spectrum on the
incident (left) side of the junction, while the parts (b) to (f)
show spectra on the transmitted (right) side for $V_0$ as
indicated. \label{bands2} }
\end{figure}

Consider now the second situation, in which the
RSOC parameter is dependent on electric field, and
let us assume the simplest case where $\lambda_1=0$. In
Fig.~\ref{bands2} we show the band structure of graphene on the
incident and transmitted sides of the junction for indicated values of the
step height $V_0$. Since the RSOC depends on the
gate voltages (and therefore on the Fermi level and $V_0$), for
each situation the RSOC parameter is adjusted according to the
formula~(\ref{alpha}). This dependence can also lead to sign reversal
of the RSOC parameter at the potential step.
To emphasize this effect, we assumed $\alpha^{-1}$ significantly
larger than currently available data for graphene~\cite{Gmitra2009}. But one may
expect, that this parameter can be increased by means of an appropriate
substrate and/or cover layers.

\begin{figure}
\includegraphics{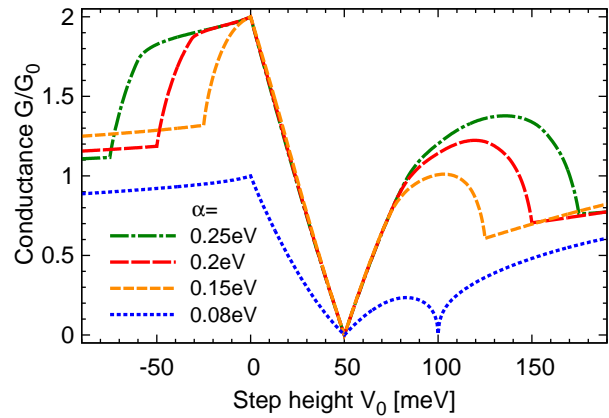}
\caption{Normalized conductance versus potential step height $V_0$
for $\lambda_1=0$, $E_F=50$ meV and indicated values of the
parameter $\alpha$.\label{Ge2}}
\end{figure}

In Fig.~\ref{Ge2} we present how the conductance $G$ depends on the
potential step height $V_0$ for different values of $\alpha$. The
first three curves (corresponding to larger values of~$\alpha$)
show similar behaviour, since in all these cases the incident
carriers come from both conduction subbands. First, there is a
minimum at $V_0=50$~meV, corresponding to vanishing density of
states on the transmitted side (RSOC vanishes as
well, see Fig.~\ref{bands2}(d)). When $V_0$ increases from this value, the
conductance grows until it reaches a~maximum value larger than
$G/G_0=1$, and then it suddenly drops. This drop occurs when one
of the subbands on the transmitted side ceases taking part in
conduction (transition from (e) to (f) in Fig.~\ref{bands2}). When
$V_0$ changes from $V_0=50$~meV towards lower values, the
normalized conductance grows until it reaches a~maximum value,
$G/G_0=2$, which is located at $V_0=0$ (no potential step, see (a)
and (b) in Fig.~\ref{bands2}). When $V_0$ goes further to negative values,
there is a~drop in conductance of the same origin as the one
discussed above.

The curve for $\alpha=0.08$ eV in Fig.~\ref{Ge2}
corresponds to the case where the incident carriers come from one
subband only. Therefore, the conductance is always equal to or
smaller than $G_0$, $G/G_0\le1$. As before, there is a~maximum at
$V_0=0$ and a~minimum at $V_0=50$ meV. Additionally, an anomalous
minimum appears at $V_0=100$ meV. This anomalous minimum is of the same nature as the one discussed in the case where $\alpha^{-1}=0$.

In summary, we have considered the charge transport through
\mbox{$p$-$n$} junctions in graphene with nonuniform RSOC parameter. The nonuniformity was either due to
asymmetric location of adjacent atomic planes (on top in one part
of the junction and at bottom in the other part), or due to a
strong dependence of the RSOC parameter on electric field (and
thus on the gate voltage used to generate carriers). We have especially focused on the situation when the RSOC
parameter changes sign at the potential step. We have found a
suppression of the conductance at characteristic points, which is a
consequence of the inverted band splitting by RSOC.

\textbf{Acknowledgements} The research has been carried out under
the project co-financed by European Union from European Regional
Fund within Operational Programme Innovative Economy.


\begin{thebibliography}{25}%
\makeatletter
\providecommand \@ifxundefined [1]{%
 \@ifx{#1\undefined}
}%
\providecommand \@ifnum [1]{%
 \ifnum #1\expandafter \@firstoftwo
 \else \expandafter \@secondoftwo
 \fi
}%
\providecommand \@ifx [1]{%
 \ifx #1\expandafter \@firstoftwo
 \else \expandafter \@secondoftwo
 \fi
}%
\providecommand \natexlab [1]{#1}%
\providecommand \enquote  [1]{``#1''}%
\providecommand \bibnamefont  [1]{#1}%
\providecommand \bibfnamefont [1]{#1}%
\providecommand \citenamefont [1]{#1}%
\providecommand \href@noop [0]{\@secondoftwo}%
\providecommand \href [0]{\begingroup \@sanitize@url \@href}%
\providecommand \@href[1]{\@@startlink{#1}\@@href}%
\providecommand \@@href[1]{\endgroup#1\@@endlink}%
\providecommand \@sanitize@url [0]{\catcode `\\12\catcode `\$12\catcode
  `\&12\catcode `\#12\catcode `\^12\catcode `\_12\catcode `\%12\relax}%
\providecommand \@@startlink[1]{}%
\providecommand \@@endlink[0]{}%
\providecommand \url  [0]{\begingroup\@sanitize@url \@url }%
\providecommand \@url [1]{\endgroup\@href {#1}{\urlprefix }}%
\providecommand \urlprefix  [0]{URL }%
\providecommand \Eprint [0]{\href }%
\providecommand \doibase [0]{http://dx.doi.org/}%
\providecommand \selectlanguage [0]{\@gobble}%
\providecommand \bibinfo  [0]{\@secondoftwo}%
\providecommand \bibfield  [0]{\@secondoftwo}%
\providecommand \translation [1]{[#1]}%
\providecommand \BibitemOpen [0]{}%
\providecommand \bibitemStop [0]{}%
\providecommand \bibitemNoStop [0]{.\EOS\space}%
\providecommand \EOS [0]{\spacefactor3000\relax}%
\providecommand \BibitemShut  [1]{\csname bibitem#1\endcsname}%
\let\auto@bib@innerbib\@empty
\bibitem [{\citenamefont {Novoselov}\ \emph {et~al.}(2004)\citenamefont
  {Novoselov}, \citenamefont {Geim}, \citenamefont {Morozov}, \citenamefont
  {Jiang}, \citenamefont {Zhang}, \citenamefont {Dubonos}, \citenamefont
  {Grigorieva},\ and\ \citenamefont {Firsov}}]{Novoselov2004}%
  \BibitemOpen
  \bibfield  {author} {\bibinfo {author} {\bibfnamefont {K.~S.}\ \bibnamefont
  {Novoselov}}, \bibinfo {author} {\bibfnamefont {A.~K.}\ \bibnamefont {Geim}},
  \bibinfo {author} {\bibfnamefont {S.~V.}\ \bibnamefont {Morozov}}, \bibinfo
  {author} {\bibfnamefont {D.}~\bibnamefont {Jiang}}, \bibinfo {author}
  {\bibfnamefont {Y.}~\bibnamefont {Zhang}}, \bibinfo {author} {\bibfnamefont
  {S.~V.}\ \bibnamefont {Dubonos}}, \bibinfo {author} {\bibfnamefont {I.~V.}\
  \bibnamefont {Grigorieva}}, \ and\ \bibinfo {author} {\bibfnamefont {A.~A.}\
  \bibnamefont {Firsov}},\ }\href {\doibase 10.1126/science.1102896} {\bibfield
   {journal} {\bibinfo  {journal} {Science}\ }\textbf {\bibinfo {volume}
  {306}},\ \bibinfo {pages} {666} (\bibinfo {year} {2004})}\BibitemShut
  {NoStop}%
\bibitem [{\citenamefont {Katsnelson}, \citenamefont {Novoselov},\ and\
  \citenamefont {Geim}(2006)}]{Katsnelson2006}%
  \BibitemOpen
  \bibfield  {author} {\bibinfo {author} {\bibfnamefont {M.~I.}\ \bibnamefont
  {Katsnelson}}, \bibinfo {author} {\bibfnamefont {K.~S.}\ \bibnamefont
  {Novoselov}}, \ and\ \bibinfo {author} {\bibfnamefont {A.~K.}\ \bibnamefont
  {Geim}},\ }\href {\doibase 10.1038/nphys384} {\bibfield  {journal} {\bibinfo
  {journal} {Nature Physics}\ }\textbf {\bibinfo {volume} {2}},\ \bibinfo
  {pages} {620} (\bibinfo {year} {2006})}\BibitemShut {NoStop}%
\bibitem [{\citenamefont {Lemme}\ \emph {et~al.}(2007)\citenamefont {Lemme},
  \citenamefont {Echtermeyer}, \citenamefont {Baus},\ and\ \citenamefont
  {Kurz}}]{Lemme2007}%
  \BibitemOpen
  \bibfield  {author} {\bibinfo {author} {\bibfnamefont {M.~C.}\ \bibnamefont
  {Lemme}}, \bibinfo {author} {\bibfnamefont {T.~J.}\ \bibnamefont
  {Echtermeyer}}, \bibinfo {author} {\bibfnamefont {M.}~\bibnamefont {Baus}}, \
  and\ \bibinfo {author} {\bibfnamefont {H.}~\bibnamefont {Kurz}},\ }\href
  {\doibase 10.1109/LED.2007.891668} {\bibfield  {journal} {\bibinfo  {journal}
  {Electron Device Letters, IEEE}\ }\textbf {\bibinfo {volume} {28}},\ \bibinfo
  {pages} {282} (\bibinfo {year} {2007})}\BibitemShut {NoStop}%
\bibitem [{\citenamefont {Yamakage}\ \emph {et~al.}(2009)\citenamefont
  {Yamakage}, \citenamefont {Imura}, \citenamefont {Cayssol},\ and\
  \citenamefont {Kuramoto}}]{Yamakage2009}%
  \BibitemOpen
  \bibfield  {author} {\bibinfo {author} {\bibfnamefont {A.}~\bibnamefont
  {Yamakage}}, \bibinfo {author} {\bibfnamefont {K.~I.}\ \bibnamefont {Imura}},
  \bibinfo {author} {\bibfnamefont {J.}~\bibnamefont {Cayssol}}, \ and\
  \bibinfo {author} {\bibfnamefont {Y.}~\bibnamefont {Kuramoto}},\ }\href
  {\doibase 10.1209/0295-5075/87/47005} {\bibfield  {journal} {\bibinfo
  {journal} {Europhys. Lett.}\ }\textbf {\bibinfo {volume} {87}},\ \bibinfo
  {pages} {47005} (\bibinfo {year} {2009})}\BibitemShut {NoStop}%
\bibitem [{\citenamefont {Yamakage}\ \emph {et~al.}(2011)\citenamefont
  {Yamakage}, \citenamefont {Imura}, \citenamefont {Cayssol},\ and\
  \citenamefont {Kuramoto}}]{Yamakage2011}%
  \BibitemOpen
  \bibfield  {author} {\bibinfo {author} {\bibfnamefont {A.}~\bibnamefont
  {Yamakage}}, \bibinfo {author} {\bibfnamefont {K.-I.}\ \bibnamefont {Imura}},
  \bibinfo {author} {\bibfnamefont {J.}~\bibnamefont {Cayssol}}, \ and\
  \bibinfo {author} {\bibfnamefont {Y.}~\bibnamefont {Kuramoto}},\ }\href
  {\doibase 10.1103/PhysRevB.83.125401} {\bibfield  {journal} {\bibinfo
  {journal} {Phys. Rev. B}\ }\textbf {\bibinfo {volume} {83}},\ \bibinfo
  {pages} {125401} (\bibinfo {year} {2011})}\BibitemShut {NoStop}%
\bibitem [{\citenamefont {Bai}\ \emph {et~al.}(2010)\citenamefont {Bai},
  \citenamefont {Wang}, \citenamefont {Tian},\ and\ \citenamefont
  {Yang}}]{Bai2010a}%
  \BibitemOpen
  \bibfield  {author} {\bibinfo {author} {\bibfnamefont {C.}~\bibnamefont
  {Bai}}, \bibinfo {author} {\bibfnamefont {J.}~\bibnamefont {Wang}}, \bibinfo
  {author} {\bibfnamefont {J.}~\bibnamefont {Tian}}, \ and\ \bibinfo {author}
  {\bibfnamefont {Y.}~\bibnamefont {Yang}},\ }\href {\doibase DOI:
  10.1016/j.physe.2010.07.011} {\bibfield  {journal} {\bibinfo  {journal}
  {Physica E}\ }\textbf {\bibinfo {volume} {43}},\ \bibinfo {pages} {207}
  (\bibinfo {year} {2010})}\BibitemShut {NoStop}%
\bibitem [{\citenamefont {Kane}\ and\ \citenamefont {Mele}(2005)}]{Kane2005}%
  \BibitemOpen
  \bibfield  {author} {\bibinfo {author} {\bibfnamefont {C.~L.}\ \bibnamefont
  {Kane}}\ and\ \bibinfo {author} {\bibfnamefont {E.~J.}\ \bibnamefont
  {Mele}},\ }\href {\doibase 10.1103/PhysRevLett.95.226801} {\bibfield
  {journal} {\bibinfo  {journal} {Phys. Rev. Lett.}\ }\textbf {\bibinfo
  {volume} {95}},\ \bibinfo {pages} {226801} (\bibinfo {year}
  {2005})}\BibitemShut {NoStop}%
\bibitem [{\citenamefont {Huertas-Hernando}, \citenamefont {Guinea},\ and\
  \citenamefont {Brataas}(2006)}]{Huertas-Hernando2006}%
  \BibitemOpen
  \bibfield  {author} {\bibinfo {author} {\bibfnamefont {D.}~\bibnamefont
  {Huertas-Hernando}}, \bibinfo {author} {\bibfnamefont {F.}~\bibnamefont
  {Guinea}}, \ and\ \bibinfo {author} {\bibfnamefont {A.}~\bibnamefont
  {Brataas}},\ }\href {\doibase 10.1103/PhysRevB.74.155426} {\bibfield
  {journal} {\bibinfo  {journal} {Phys. Rev. B}\ }\textbf {\bibinfo {volume}
  {74}},\ \bibinfo {pages} {155426} (\bibinfo {year} {2006})}\BibitemShut
  {NoStop}%
\bibitem [{\citenamefont {Min}\ \emph {et~al.}(2006)\citenamefont {Min},
  \citenamefont {Hill}, \citenamefont {Sinitsyn}, \citenamefont {Sahu},
  \citenamefont {Kleinman},\ and\ \citenamefont {MacDonald}}]{Min2006}%
  \BibitemOpen
  \bibfield  {author} {\bibinfo {author} {\bibfnamefont {H.}~\bibnamefont
  {Min}}, \bibinfo {author} {\bibfnamefont {J.~E.}\ \bibnamefont {Hill}},
  \bibinfo {author} {\bibfnamefont {N.~A.}\ \bibnamefont {Sinitsyn}}, \bibinfo
  {author} {\bibfnamefont {B.~R.}\ \bibnamefont {Sahu}}, \bibinfo {author}
  {\bibfnamefont {L.}~\bibnamefont {Kleinman}}, \ and\ \bibinfo {author}
  {\bibfnamefont {A.~H.}\ \bibnamefont {MacDonald}},\ }\href {\doibase
  10.1103/PhysRevB.74.165310} {\bibfield  {journal} {\bibinfo  {journal} {Phys.
  Rev. B}\ }\textbf {\bibinfo {volume} {74}},\ \bibinfo {pages} {165310}
  (\bibinfo {year} {2006})}\BibitemShut {NoStop}%
\bibitem [{\citenamefont {Gmitra}\ \emph {et~al.}(2009)\citenamefont {Gmitra},
  \citenamefont {Konschuh}, \citenamefont {Ertler}, \citenamefont
  {Ambrosch-Draxl},\ and\ \citenamefont {Fabian}}]{Gmitra2009}%
  \BibitemOpen
  \bibfield  {author} {\bibinfo {author} {\bibfnamefont {M.}~\bibnamefont
  {Gmitra}}, \bibinfo {author} {\bibfnamefont {S.}~\bibnamefont {Konschuh}},
  \bibinfo {author} {\bibfnamefont {C.}~\bibnamefont {Ertler}}, \bibinfo
  {author} {\bibfnamefont {C.}~\bibnamefont {Ambrosch-Draxl}}, \ and\ \bibinfo
  {author} {\bibfnamefont {J.}~\bibnamefont {Fabian}},\ }\href {\doibase
  10.1103/PhysRevB.80.235431} {\bibfield  {journal} {\bibinfo  {journal} {Phys.
  Rev. B}\ }\textbf {\bibinfo {volume} {80}},\ \bibinfo {pages} {235431}
  (\bibinfo {year} {2009})}\BibitemShut {NoStop}%
\bibitem [{\citenamefont {Wallace}(1947)}]{Wallace1947}%
  \BibitemOpen
  \bibfield  {author} {\bibinfo {author} {\bibfnamefont {P.~R.}\ \bibnamefont
  {Wallace}},\ }\href {\doibase 10.1103/PhysRev.71.622} {\bibfield  {journal}
  {\bibinfo  {journal} {Physical Review}\ }\textbf {\bibinfo {volume} {71}},\
  \bibinfo {pages} {622} (\bibinfo {year} {1947})}\BibitemShut {NoStop}%
\bibitem [{\citenamefont {Varykhalov}\ \emph {et~al.}(2008)\citenamefont
  {Varykhalov}, \citenamefont {S\'anchez-Barriga}, \citenamefont {Shikin},
  \citenamefont {Biswas}, \citenamefont {Vescovo}, \citenamefont {Rybkin},
  \citenamefont {Marchenko},\ and\ \citenamefont {Rader}}]{Varykhalov2008}%
  \BibitemOpen
  \bibfield  {author} {\bibinfo {author} {\bibfnamefont {A.}~\bibnamefont
  {Varykhalov}}, \bibinfo {author} {\bibfnamefont {J.}~\bibnamefont
  {S\'anchez-Barriga}}, \bibinfo {author} {\bibfnamefont {A.~M.}\ \bibnamefont
  {Shikin}}, \bibinfo {author} {\bibfnamefont {C.}~\bibnamefont {Biswas}},
  \bibinfo {author} {\bibfnamefont {E.}~\bibnamefont {Vescovo}}, \bibinfo
  {author} {\bibfnamefont {A.}~\bibnamefont {Rybkin}}, \bibinfo {author}
  {\bibfnamefont {D.}~\bibnamefont {Marchenko}}, \ and\ \bibinfo {author}
  {\bibfnamefont {O.}~\bibnamefont {Rader}},\ }\href {\doibase
  10.1103/PhysRevLett.101.157601} {\bibfield  {journal} {\bibinfo  {journal}
  {Phys. Rev. Lett.}\ }\textbf {\bibinfo {volume} {101}},\ \bibinfo {pages}
  {157601} (\bibinfo {year} {2008})}\BibitemShut {NoStop}%
\bibitem [{\citenamefont {S\'anchez-Barriga}\ \emph {et~al.}(2010)\citenamefont
  {S\'anchez-Barriga}, \citenamefont {Varykhalov}, \citenamefont {Scholz},
  \citenamefont {Rader}, \citenamefont {Marchenko}, \citenamefont {Rybkin},
  \citenamefont {Shikin},\ and\ \citenamefont {Vescovo}}]{Sanchez-Barriga2010}%
  \BibitemOpen
  \bibfield  {author} {\bibinfo {author} {\bibfnamefont {J.}~\bibnamefont
  {S\'anchez-Barriga}}, \bibinfo {author} {\bibfnamefont {A.}~\bibnamefont
  {Varykhalov}}, \bibinfo {author} {\bibfnamefont {M.}~\bibnamefont {Scholz}},
  \bibinfo {author} {\bibfnamefont {O.}~\bibnamefont {Rader}}, \bibinfo
  {author} {\bibfnamefont {D.}~\bibnamefont {Marchenko}}, \bibinfo {author}
  {\bibfnamefont {A.}~\bibnamefont {Rybkin}}, \bibinfo {author} {\bibfnamefont
  {A.}~\bibnamefont {Shikin}}, \ and\ \bibinfo {author} {\bibfnamefont
  {E.}~\bibnamefont {Vescovo}},\ }\href {\doibase DOI:
  10.1016/j.diamond.2010.01.047} {\bibfield  {journal} {\bibinfo  {journal}
  {Diamond Relat. Mater.}\ }\textbf {\bibinfo {volume} {19}},\ \bibinfo {pages}
  {734 } (\bibinfo {year} {2010})}\BibitemShut {NoStop}%
\bibitem [{\citenamefont {Rashba}(2009)}]{Rashba2009}%
  \BibitemOpen
  \bibfield  {author} {\bibinfo {author} {\bibfnamefont {E.~I.}\ \bibnamefont
  {Rashba}},\ }\href {\doibase 10.1103/PhysRevB.79.161409} {\bibfield
  {journal} {\bibinfo  {journal} {Phys. Rev. B}\ }\textbf {\bibinfo {volume}
  {79}},\ \bibinfo {pages} {161409} (\bibinfo {year} {2009})}\BibitemShut
  {NoStop}%
\bibitem [{\citenamefont {Li}\ \emph {et~al.}(2011)\citenamefont {Li},
  \citenamefont {Yang}, \citenamefont {Qiao}, \citenamefont {Hu},\ and\
  \citenamefont {Wu}}]{Li2011}%
  \BibitemOpen
  \bibfield  {author} {\bibinfo {author} {\bibfnamefont {Z.~Y.}\ \bibnamefont
  {Li}}, \bibinfo {author} {\bibfnamefont {Z.~Q.}\ \bibnamefont {Yang}},
  \bibinfo {author} {\bibfnamefont {S.}~\bibnamefont {Qiao}}, \bibinfo {author}
  {\bibfnamefont {J.}~\bibnamefont {Hu}}, \ and\ \bibinfo {author}
  {\bibfnamefont {R.~Q.}\ \bibnamefont {Wu}},\ }\href
  {http://stacks.iop.org/0953-8984/23/i=22/a=225502} {\bibfield  {journal}
  {\bibinfo  {journal} {J. Phys.: Condens. Matter}\ }\textbf {\bibinfo {volume}
  {23}},\ \bibinfo {pages} {225502} (\bibinfo {year} {2011})}\BibitemShut
  {NoStop}%
\bibitem [{\citenamefont {Castro~Neto}\ and\ \citenamefont
  {Guinea}(2009)}]{CastroNeto2009}%
  \BibitemOpen
  \bibfield  {author} {\bibinfo {author} {\bibfnamefont {A.~H.}\ \bibnamefont
  {Castro~Neto}}\ and\ \bibinfo {author} {\bibfnamefont {F.}~\bibnamefont
  {Guinea}},\ }\href {\doibase 10.1103/PhysRevLett.103.026804} {\bibfield
  {journal} {\bibinfo  {journal} {Phys. Rev. Lett.}\ }\textbf {\bibinfo
  {volume} {103}},\ \bibinfo {pages} {026804} (\bibinfo {year}
  {2009})}\BibitemShut {NoStop}%
\bibitem [{\citenamefont {Ertler}\ \emph {et~al.}(2009)\citenamefont {Ertler},
  \citenamefont {Konschuh}, \citenamefont {Gmitra},\ and\ \citenamefont
  {Fabian}}]{Ertler2009}%
  \BibitemOpen
  \bibfield  {author} {\bibinfo {author} {\bibfnamefont {C.}~\bibnamefont
  {Ertler}}, \bibinfo {author} {\bibfnamefont {S.}~\bibnamefont {Konschuh}},
  \bibinfo {author} {\bibfnamefont {M.}~\bibnamefont {Gmitra}}, \ and\ \bibinfo
  {author} {\bibfnamefont {J.}~\bibnamefont {Fabian}},\ }\href {\doibase
  10.1103/PhysRevB.80.041405} {\bibfield  {journal} {\bibinfo  {journal} {Phys.
  Rev. B}\ }\textbf {\bibinfo {volume} {80}},\ \bibinfo {pages} {041405}
  (\bibinfo {year} {2009})}\BibitemShut {NoStop}%
\bibitem [{\citenamefont {Dedkov}\ \emph {et~al.}(2008)\citenamefont {Dedkov},
  \citenamefont {Fonin}, \citenamefont {R\"udiger},\ and\ \citenamefont
  {Laubschat}}]{Dedkov2008}%
  \BibitemOpen
  \bibfield  {author} {\bibinfo {author} {\bibfnamefont {Y.~S.}\ \bibnamefont
  {Dedkov}}, \bibinfo {author} {\bibfnamefont {M.}~\bibnamefont {Fonin}},
  \bibinfo {author} {\bibfnamefont {U.}~\bibnamefont {R\"udiger}}, \ and\
  \bibinfo {author} {\bibfnamefont {C.}~\bibnamefont {Laubschat}},\ }\href
  {\doibase 10.1103/PhysRevLett.100.107602} {\bibfield  {journal} {\bibinfo
  {journal} {Phys. Rev. Lett.}\ }\textbf {\bibinfo {volume} {100}},\ \bibinfo
  {pages} {107602} (\bibinfo {year} {2008})}\BibitemShut {NoStop}%
\end{thebibliography}
\end{document}